\documentclass[12pt,preprint]{aastex}

\begin{document}

\title{HST/NICMOS Imaging Survey of the Ophiuchus (Lynds 1688) Cluster}

\author{Lori E. Allen, Philip C. Myers, James Di Francesco\altaffilmark{1}}
\affil{Harvard--Smithsonian Center for Astrophysics, 60 Garden Street, MS42,  
    Cambridge, MA 02138}

\author{Robert Mathieu}
\affil{Department of Astronomy, University of Wisconsin, Madison, WI 53706}

\author{Hua Chen\altaffilmark{2} and Erick Young}
\affil{Steward Observatory, University of Arizona, Tucson AZ 85721}

\altaffiltext{1}{Currently at the Radio Astronomy Laboratory, 601 Campbell Hall, University of California at Berkeley, Berkeley, CA 94720-3411}
\altaffiltext{2}{Currently at Nortel Networks, 2305 Mission College Blvd, Santa Clara, CA 95054}


\begin{abstract}
We present a catalogue of near-infrared photometry of 
young stars associated with the   
Ophiuchus molecular cloud, based on observations made with the 
Hubble Space Telescope NICMOS-3 camera at 1.1~$\mu$m and 1.6~$\mu$m. 
Our survey covers 0.02 square degrees centered on the dense molecular 
cores in Lynds 1688.
We detect 165 sources at 1.6~$\mu$m and 
65 sources at 1.1~$\mu$m, within our estimated completeness limits of 21.0 mag and 
21.5 mag, respectively. An analysis of the cloud extinction, based on existing 
molecular line maps, suggests that most of the sources lying within the 40 $\rm{A_V}$ 
extinction contour of the cloud are probable cloud members. Approximately half 
(58/108) of these sources are previously unpublished. 

The faint embedded sources revealed by these observations are spatially concentrated 
in three regions of high stellar space density (N$> 10^4$ stars pc$^{-3}$).  While 
the spatial distribution of these sources reflects that of the brighter, well--known 
population of young stars in Ophiuchus, it is distinctly different from the distribution 
of cool concentrations seen in the submillimeter.  
Seven new brown dwarf candidates are identified, based on their infrared colors and 
their projected locations on high column-density regions 
of the molecular cloud. 
Eight new candidate binary and five new candidate triple systems, having separations 
between 0\farcs2 to 10\arcsec~ (29 to 1450 AU) are reported.  
The spatial resolution and sensitivity of these observations reveal five apparent 
disk/envelope systems seen 
via scattered light, and four nebulous objects with complex morphologies. 

\end{abstract}




\section{Introduction}

Among molecular clouds within 200 pc of the Sun, the Ophiuchus cloud 
offers the best opportunity 
to study the formation of a stellar cluster at close range. 
At a distance of 145 pc \citep{deZ99}, 
it contains a high density of low-mass young stellar objects 
(YSOs) deeply embedded in the cloud material, as evidenced by 
the extremely high gas and dust column densities measured there  
\citep{lww90,lor86,wil83}. 
 
Most of the stellar content of the Ophiuchus cloud 
is obscured at 
visible wavelengths.  
However, early infrared observations revealed  
a rich embedded stellar 
population \citep{gss,vrb73,
faz76,el78,wil83}.  More recent observations using infrared arrays combined with molecular line maps have shown 
that these stars are concentrated in three dense molecular cores within the 
cloud \citep{gy,com93,sks,bklt}.  Age estimates based on infrared 
spectroscopic studies indicate that the stars are very young, with ages 
less than 10$^6$ yr \citep{gm95,wgm99,lr99}. 

Recent evidence suggests that most stars form in clusters \citep{carp00}, 
but there is
relatively little theoretical guidance available to indicate how clusters
develop \citep{myers2000,adams2001,meyer}.  The hierarchical structure of some
groups has been said to arise from the hierarchical nature of condensations
formed by turbulent processes (e.g. Klessen et al. 2000), but tests of
cluster formation models require better knowledge of cluster spatial
structure, and its temporal development. The youngest clusters are the best
regions to study for this purpose because their stars have had relatively
little time to move from their formation sites, and the spatial structure
of the youngest members is essentially their spatial structure at birth.

In this contribution, we present the results of an infrared imaging survey 
using NICMOS on the Hubble Space Telescope (HST). 
The superior sensitivity and spatial resolution 
provided by HST were used to i) obtain a more complete census of the young 
stellar population in the dense star-forming regions of the cloud, 
ii) determine the degree of clustering and the binary and multiple star 
frequency, iii) identify candidate brown dwarfs, and iv) resolve the 
morphologies of deeply embedded extended sources. 

Details of the observations and data analysis are described in \S~\ref{obs}. 
Positions and photometry of all detected sources are presented in 
Table 2, along with cross-correlations with other designations 
from the literature. 
In \S~\ref{resu}, we present results on the clustering and multiplicity 
of our sample, and a search for new brown dwarf candidates. A 
detailed description of extended sources is provided in \S~\ref{fuzz}. 
A list of binary and multiple 
systems is presented in Table 3, and new candidate brown dwarfs are 
noted in Table 2. These results are summarized in \S~\ref{summ}.

\section{Observations and Data Analysis}\label{obs}

\subsection{Observations}

Observations were obtained with HST/NICMOS Camera 3 during the 
June 1998 campaign. Camera 3 has a pixel scale of 0\farcs20 pixel$^{-1}$   
and a field--of--view of $51\arcsec\times51\arcsec$.~   
Images were made through the F110W and F160W filters, with integration times of 
39.95 seconds and 31.96 seconds, respectively.  Thirteen target positions
were chosen to correspond with both high surface density of previously 
known YSOs and high gas column density, and are listed in Table~\ref{tbl-1} 
(see also Figure 1). 
Each was observed in a $3\times3$ spiral pattern with a 48\arcsec~ dither,  
resulting in 13 $2\farcm45\times2\farcm45$ mosaics. 
Each field was imaged twice 
in each filter, with a 3\arcsec~ offset between image pairs to aid in identification and 
removal of bad pixels and cosmic rays.


\subsection{Image Processing and Calibration}

Uncalibrated images were processed using the IRAF/STSDAS package CALNICA, and the 
latest reference files provided by the Space Telescope Science Institute (STScI).  
Because Camera 3 suffers from a DC offset or bias problem 
(a.k.a. the ``pedestal effect''), corrections were made using a program in IDL (PEDTHERM)  provided by L. Bergeron of STScI. In regions dominated by bright extended emission,
no pedestal corrections were made.
For each mosaic, a sky frame was constructed of median filtered images, using 
only those fields which did not contain bright extended emission. 
Masks were constructed to prevent stars from contributing to both the 
pedestal and sky determinations.   
Sky-subtracted image pairs were cross-correlated to determine spatial shifts, and combined 
using the IRAF/STSDAS package DRIZZLE \citep{fru97}, which re-samples the image pairs onto 
a common pixel grid. Images containing 
no apparent sources were combined with DRIZZLE using the prescribed telescope dither. 

Reduced images were flux calibrated using recently determined photometric scale factors, 
$ 2.873\times10^{-6}$ Jy $\rm{(ADU/sec)^{-1}}$ at 1.1~$\mu$m and $2.776\times10^{-6}$ Jy $\rm{(ADU/sec)^{-1}}$ at 1.6~$\mu$m \citep{rieke99}.
Corresponding zero points were calculated on the Vega system, assuming 
zero magnitude flux densities at 1775 and 1083 Jy and effective wavelengths of 
1.104~$\mu$m and 1.593~$\mu$m 
for F110W and F160W respectively \citep{rieke99}.

\subsection{Photometry and Astrometry} \label{phot}

Magnitudes were measured using aperture photometry routines in IRAF/DIGIPHOT, and are 
reported in Table 2. 
For point sources we used an aperture of radius 0\farcs5~ (5 pixels in our drizzled images). 
A larger aperture radius of 3\arcsec~ was used to measure the flux from 
extended sources, as noted in Table 2. 
By adding artificial stars to our data, 
we estimate that our photometry is 90\% complete to 21.0 mag at F160W and 21.5 mag at F110W.   
Within those limits, we detect 165 sources at F160W and 65 sources at F110W.
A census of near-infrared surveys in the literature 
\citep{lr99, bklt, sks, com93, gy} 
indicates that approximately two thirds of the sources  
listed in Table 2 are previously unreported.

Enough overlap exists between our survey and previous studies that we can derive 
transformations between the F110W and F160W magnitudes and a standard 
ground-based system. Comparing our photometry with that of Barsony et al. (1997) for 
21 sources, we find the following linear relations: 
${\rm m(F110W)=(1.07\pm0.001)\times J - (0.44\pm0.012)}$,  and  ${\rm m(F160W)=(1.02\pm0.001)\times H + (0.03\pm0.007)}$, 
where J and H are on the CIT system. 
These relations were used to convert our F110W and F160W magnitudes to J and H 
magnitudes, allowing their placement in a color-magnitude diagram 
( see Figure~\ref{fig:cm}, discussed in \S~\ref{bd}). 
 
Because no guide stars were visible in the highly extincted target fields, 
observations were unguided, 
leading to some drift in the field centers 
(of order a few to 20\arcsec~ throughout a 3x3 map). 
This drift, coupled with the fact that many 
frames have no or    
very few sources, complicated the determination of accurate astrometric solutions 
with these data. 
Instead, the coordinate system of Barsony et al. (1997) was adopted, 
since that survey covers a large 
fraction of the cloud and has been extensively cross-correlated with previous studies.
Barsony et al. estimated 1$\sigma$ uncertainties of 1\farcs2~ in their absolute positions.

\section{Results}\label{resu}

\subsection{Cloud membership and background stars}\label{bground}

Given the sensitivity of our survey, we can expect to detect some   
background stars through the molecular cloud, so some means of 
distinguishing cloud members from background sources would be helpful. 
Unfortunately, with photometry in only 2 bands we cannot use colors 
to distinguish 
reddened background field stars from non--nebulous  
pre--main sequence stars embedded within the cloud.
However, we can estimate the background 
by considering the  
observed distribution of objects with respect to the column density 
of molecular gas, and the expected 
source counts in this part of the sky from a model of the Galaxy. 

In Figure~\ref{fig:avmap}, we reproduce the extinction map made by 
Wilking \& Lada (1983), based on observations 
of C$^{18}$O (J=1--0) and $^{12}$CO (J=1--0). 
On it we plot the locations of all 165 sources detected at F160W. 
Our HST mosaics are coincident with high $\rm{A_V}$ regions, with the exception 
of some fields on the eastern sides of cores A, B, and F.  These fields 
coincide with a steep gradient in the cloud column density, 
and in them large numbers of stars are seen. As the extinction increases 
toward the center of the cloud, the number of stars decreases. 
Clearly, a significant fraction 
of detected sources on the edges of the cloud cores must be background 
stars. 

The infrared Galactic model of Wainscoat et al. (1992) predicts the 
existence of approximately 140 stars arcmin$^{-2}$ 
brighter than our detection limit in the direction of the  
Ophiuchus cloud, assuming $\rm{A_V}=0$. 
However, when the model prediction is convolved with the $\rm{A_V}$ map, 
the resulting background is substantially smaller. 
This was done by calculating the expected background for each pixel in the $\rm{A_V}$ map, 
then summing over all pixels 
to obtain the total number of expected background stars  
as a function of H magnitude. 

In Figure~\ref{fig:hlum} we plot the observed H-band distribution of stars  
(using the conversion from magnitude at F160W to magnitude at H given 
in Section~\ref{phot}). The distribution predicted by 
Wainscoat's model and reddened with our $\rm{A_V}$ map (assuming 
$\rm{A_H}$ = 0.155$\rm{A_V}$, \cite{coh81}) is shown as a heavy dotted line.  
It appears that the 
background is small, becoming dominant only for H$>$20 mag. 
For comparison, we also plot the expected background as seen through a 
uniform extinction cloud of $\rm{A_V} =$ 35, 40, and 45 mag. 
The background modelled with our extinction map agrees closely 
with that for a uniform extinction of 40 mag. For this reason, 
we shall assume that all stars lying outside the $\rm{A_V}=$40 mag. 
 contour in 
Figure~\ref{fig:avmap} are background. 
Within the 40 mag.  
contour, the probable number of background stars given by the  
Wainscoat model ranges from approximately 0.50 stars 
arcmin$^{-2}$ (for $\rm{A_V}$=40 mag) to 0.02 stars 
arcmin$^{-2}$ (for $\rm{A_V}$=80 mag). 
Of the 165 sources detected in our survey, 
108 sources lie within the 40 mag contour, and 58 of these are new detections, 
too faint to have been detected in previous surveys. 
In the discussions of clustering, 
multiplicity, and brown dwarfs which follow, only these 108 sources will be 
considered. 


\subsection{The spatial distribution of young stars in the Ophiuchus cloud}\label{cluster}

Ground--based imaging surveys at 2~$\mu$m 
revealed that the young stars in the Ophiuchus cloud are clustered in 3--4  
main groups associated with dense cores of molecular gas  
\citep{lr99,bklt,sks,com93,gy,bar89,rab89}.
Our HST survey targeted these peaks in the YSO surface density distribution, allowing us 
to examine structure  
within these stellar concentrations to a greater depth and with higher 
resolution than before. 

The highest concentration of sources 
is in core A, where 23 stars are detected at F160W within an area   
$0.08\times0.05$ pc in size (1\farcm9 x 1\farcm2~, centered on HST position 2 in 
Table 1). 
The extinction in this area is $\rm{A_V}=45-80$ mag, and averages 
60 mag, leading to an estimated 
background contribution of 0.3$\pm$2 stars. The YSO surface density in the area, 
excluding background objects, is then $\sim5\times10^3$ stars pc$^{-2}$.
Assuming a range of depths from half 
to twice the area width, we estimate stellar volume densities of 
$\sim4-10\times10^4$ stars pc$^{-3}$, 
comparable to the density of $2\times10^4$ stars pc$^{-3}$ determined for the core of the 
Orion nebula cluster\footnote{It is worth noting that, while these space densities are similar, the ONC core
occupies a volume about 10 times larger than the one we describe in Oph, and contains
hundreds of stars.}
(Hillenbrand \& Hartmann 1998). 
Other peaks in the stellar density are found in core B (HST positions 9 and 10) 
and core E (HST positions 12 and 13). 
Volume densities there are slightly less, ranging from $0.6-5\times 10^4$ stars pc$^{-3}$.

Inverting the volume density yields a mean spacing between stars in the core A 
peak of $0.02 - 0.03$ pc, or $\sim4000-5000$ AU, and $\sim5600-11000$ AU in the 
B and E cores. 
This is similar to the 6000 AU ``fragmentation'' scale identified by 
Motte et al. (1998) in their analysis of 1.3 mm continuum dust clumps, 
which are concentrated in the  
same three regions as the infrared sources reported here. 

In fact, the embedded stars in Ophiuchus and the dust clumps detected at 
1.3 mm have similar, but not identical, spatial distributions. 
The clumps show the same sub-clustering shown by the stars, as seen in 
Figure~\ref{fig:space}, where the positions of the 
starless clumps from Motte et al. (1998; open squares) 
have been plotted along with the positions of the stars (filled circles). 
Interestingly, the clumps are more strongly associated with the 
highest extinction regions of the cloud, whereas the stars appear 
to cluster around the edges of these regions. 

This similar spatial relation of the stars and the millimeter continuum clumps 
in each of the three subgroups of the Ophiuchus core suggests that we are 
observing a real effect with the same explanation (origin?) in each 
subgroup. Some of the millimeter continuum clumps could be highly 
extincted YSOs, but this seems unlikely because 
most of the clumps 
are extended, some have associated compact 
molecular line emission, and 
the necessary extinction, $\rm{A_V}>1000$ mag, 
is in all other known cases accompanied by emission at far- and mid-infrared 
wavelengths (cf. Ladd et al. 1991).  Adopting the interpretation of Motte 
et al. (1998) that the millimeter continuum clumps are prestellar, a more likely 
explanation is that in each subgroup, the region of highest extinction 
is a ``starless core'' which has not yet formed stars, but has formed numerous  protostellar condensations with relative spacing (4000-5000 AU) 
similar to that of the 
surrounding young stars. In this picture, the similar spacing of prestellar 
and young stellar objects suggests that the young stellar objects have neither concentrated nor dispersed significantly since their formation. This evidence 
for spatial segregation by age may offer important clues to how stars form 
in clusters. 

\subsection{Multiplicity and clustering}\label{multi}

To search for apparent binary pairs and higher-order 
multiple systems, we performed both a visual examination of all 
images and a nearest neighbor calculation for all detected sources. 
However in order to decrease the likelihood of mistaking 
background stars for companions, 
we restricted the sample to the 108 stars  
projected within the $\rm{A_V} \ge$40 mag contour of the cloud 
(as discussed in \S~\ref{bground}).

For relatively faint stars (m(F160W)$\ge$18), our survey is sensitive to 
separations comparable to the resolution of the array 
(0\farcs2 pixel$^{-1}$), although due to the complex NICMOS PSF, 
our detection of faint companions within $\theta < 0.5 $\arcsec~ 
of bright stars may be incomplete.
To guard against bias, we took our estimated limiting magnitude for 
faint companions (m(F160W)$\ge$18) and imposed this limit on the sample,
further restricting our sample to 62 stars. 
The maximum separation considered in our search was 10\arcsec (1450 AU), 
chosen to coincide with previous searches for pre-main sequence binaries. 
Within the separation range of 0\farcs2 to 10\arcsec, we detected   
seven apparent binary pairs and six apparent triple systems. They are 
listed in Table 3, and shown in
Figure~\ref{fig:multis}. 
The triple systems pictured in panels i, j, and k in Figure~\ref{fig:multis}, 
are unusual in that they appear to be non-hierarchical. The other triple systems 
reported here (panels l and m) are organized in the usual way, having a close pair 
with a widely separated third member.

The multiplicity fraction, defined as the ratio of the 
number of binary and multiple 
systems detected, to 
the number of single, binary and multiple systems observed, or    
mf=B+M/(S+B+M), is then 0.30$\pm$0.08.  
Several previous multiplicity studies have included targets in Lynds 1688,  
each 
one sensitive to a specific range of companion separations and limiting magnitudes. 
In the lunar occultation experiment of Simon et al. (1995), observations were 
sensitive to stars 
brighter than K$\sim$11 and binary  
separations of 0\farcs005~ to 10\arcsec.~ Observing 35 systems, they 
detected 10 binaries, two triples, and one quadruple, for  
a multiplicity fraction of 0.37$\pm$0.10. 

There are important differences  
between Simon et al. (1995; hereafter S95) and this study which should be considered 
when comparing their results. 
First, the minimum separation detectable by S95 was  
0\farcs005, whereas ours is $\sim$0\farcs2. 
The limiting magnitude of the samples also differ:   
$\rm{K_{lim}}\sim12$ and $\rm{H_{lim}}=18$ in S95 and this study, 
respectively.  
For projected separations between 0\farcs2~ and 10\arcsec~ only, 
S95 obtain mf$\sim0.26\pm0.08$. 
When we compare this with our mf of $0.30\pm0.08$ for a deeper limiting magnitude, 
we conclude that the two results do not differ significantly. 
Within the uncertainties, 
the multiplicity fractions of S95 and this work are in agreement. 

The relationship between the separation of binary pairs and 
clustering on a larger scale has been explored in a number of 
nearby star forming regions, including Ophiuchus \citep{ghez93, reizen93, 
sks, nak98}.
Of particular interest is the mean surface density of companions 
(MSDC), which Larson (1995) applied to young stars
in the Taurus-Auriga star-forming region, finding clustering 
on two distinct scales. For large-scale clustering 
the MSDC was found to have a power-law slope of $\approx$ $-0.6$, 
and for small separations a slope of $\approx$ $-2$ was found. 
The ``break point'' between large and small scale slopes was determined 
to be $\approx$ $0.04$ pc.  Simon (1997; hereafter S97) computed the MSDC for Ophiuchus, 
Taurus, and the Trapezium, finding similar power laws for each.  

We computed the MSDC for the 
108 stars projected within the $\rm{A_V} \ge$40 mag contour of Figure 2.
The result is shown in Figure 6 for two samples of different depths: a shallow 
magnitude cutoff (H$\le$14) and a deep limiting magnitude (H$\le$21). As seen in 
Figure 6, the two samples have different break points. The shallow magnitude 
limit was selected to approximately match that of S97 (K$\le$12), and 
not surprisingly, the two MSDCs have the same slopes and break points. The deeper 
sample shows a break point on a smaller spatial scale (by an order of magnitude). 
Thus, in a given star forming region, the properties of the MSDC can vary according 
to the depth of the sample. Bate et al. (1998) cautioned against overinterpreting the 
MSDC, in part for this reason. We offer this comparison as further caution.

\subsection{New candidate brown dwarfs} \label{bd}
  
With completeness limits of 21.5 and 21.0 magnitudes at F110W and F160W respectively,
our survey is sensitive to very low-mass objects. For example, we should  
have detected stars of spectral type L4 V, through as much as 17 $\rm{A_V}$, 
assuming $\rm{M_J} = 13.2$ and $\rm{M_H} = 12.3$ \citep{jdk99}.
Such very low-mass objects are relatively ``blue'' 
in intrinsic color, typically having  $\rm{(J-H)<2}$ (Kirkpatrick et al. 1999), 
and so might be distinguishable 
from heavily reddened background stars in an infrared color-magnitude diagram. 

Using the transformations from F110W and F160W to J and H listed in \S~\ref{phot} 
and the fluxes in Table 2,  
we converted our measured magnitudes to J and H magnitudes and plotted them in the  
color-magnitude diagram shown in Figure~\ref{fig:cm}. Also plotted are 
the $3\times 10^5$ yr isochrone and the ZAMS from the pre-main sequence stellar 
evolution models of D'Antona \& Mazzitelli (1997). 
Only the 65 sources detected in both the F110W and F160W bands are shown. 

Brown dwarfs cannot be positively identified using near-IR photometry alone. 
However, those sources which 
are faint and relatively blue are more likely to be low-mass objects in the cloud 
than background stars seen through the cloud, especially if they are coincident with 
high extinction regions. 
We identified ten candidate brown dwarfs in our sample, using the 
following criteria:  
i) (J-H)$<$ 3.0, ii) J$>$15, and  
iii) position coincident with cloud extinction values of $\rm{A_V}>50$ mag. 
Spectral types have been published for 
three of these objects. One object, 162622$-$242409 is almost certainly sub-stellar, 
having been assigned 
a spectral type of M6.5 by both Wilking, Greene, \& Meyer (1999;
hereafter WGM) and Luhman \& Rieke (1999; hereafter LR), and having 
a luminosity of 
0.002-0.003 $\rm L_{\odot}$.
The mass of object 162622$-$242354 is less certain. 
WGM classify it as M8.5  
with $\rm L_{bol}\sim 0.004 L_{\odot}$, 
whereas LR adopt a spectral type   
of M6.5
and a luminosity of 0.067 $\rm L_{\odot}$. 
The third classified object, 162659$-$243556, is a star of type M4 (LR), too 
hot to be a brown dwarf. 
The remaining seven (new) brown dwarf candidates are noted in Table 2. 
They are located in core A, which contains the highest density of stars in the 
survey (\S~3.1). 
Two of the candidates, 162622$-$242409 and 162622$-$242408, are members of triple systems
(\S~\ref{multi} and Table 3). One of the candidates, 16265$-$242303, is just on the red   
edge of our color criterion, but qualifies for selection when 
photometric uncertainties are taken into account. 

There are a total of ten known brown dwarfs in the Ophiuchus cluster 
(LR, WGM) and another five sources known to occupy the transition region in the 
HR diagram between the stellar and substellar regimes (WGM).
If all of the seven new candidates identified here are indeed substellar, 
the number of brown dwarfs known in the Ophiuchus cluster 
will have increased by 50\%-70\%.
We can assess the 
implications for the stellar initial mass function 
(IMF) in the cluster.   
However, because fainter, lower-mass objects cannot be seen as deeply 
into the cloud as higher-mass stars, we restrict our analysis to an extinction-limited 
sample. In Figure~\ref{fig:cm}, there are 30 sources within an 
extinction-limited region delimited by the $3\times10^5$ yr isochrone 
and the 
$\rm{A_V}<20$ mag limit. These include our seven new candidate brown dwarfs, 
and the two confirmed brown dwarfs. 
If all seven of the new candidates are brown dwarfs, 
the total fraction of substellar 
objects in the extinction-limited sample is 9/30 = 30\%.
This is consistent with the results of LR, who argued that 
the IMF includes a large fraction of sub-stellar mass objects. 

\section{Morphologies of extended sources}\label{fuzz}

Several interesting extended objects were covered by our survey, 
including the well-known infrared sources GSS 30, YLW 15A and YLW 16A, 
as well as some catalogued  
objects whose morphologies were previously 
unresolved.   
These sources are shown in Figure~\ref{fig:plate} 
and described in detail below. 

\noindent
{\bf GSS 30 (162621$-$242306):}
This multiple source is perhaps the most studied object in Ophiuchus. 
The illuminating star GSS 30--IRS 1 (162621$-$242306) 
is a class I source of unknown spectral type. 
A K-band spectrum obtained by LR  
showed the source to be heavily veiled (r$>$2) with no 
photospheric absorption features, and with Br$\gamma$ in 
emission. 
Two other infrared sources, GSS 30--IRS 2 and GSS 30--IRS 3,
 are located within the projected bounds of the 
nebula. 

The nebula illuminated by IRS 1 has itself been the subject of 
many investigations. 
Near-infrared polarimetry images \citep{chrys97,chrys96,wein93,tam91,cas85}  
delineate a bipolar morphology coincident with the near-IR emission on 
the northeast side, and extending to the southwest side, where there 
is less infrared emission. The polarization pattern is approximately  
centrosymmetric around the illuminating source, and shows a 
``polarization disk'' orthogonal to the long axis of the 
nebula. 
Zhang et al. (1997) detected a somewhat flattened molecular core 
in C$^{18}$O and 
$^{13}$CO (J=1--0), in roughly the same orientation as the polarization 
disk.  
In the high resolution image presented in Figure~\ref{fig:plate} 
one sees the characteristic 
hourglass morphology of a bipolar nebula. The emission is dominated by 
the northeast lobe, 
a fact which has led previous 
investigators to propose that the fainter southwest lobe is obscured 
by a tilted circumstellar disk or toroid \citep{cas85, tam91, wein93, chrys97}. 

{\bf GSS 30--IRS 2 (162622$-$242254)}
is a class III source of late K/early M spectral type 
\citep{lr99, gm95}, located 20\arcsec~ to the northeast 
of GSS 30--IRS 1. Its relatively advanced evolutionary class 
indicates it is probably unrelated and   
slightly foreground to the nebula. 

{\bf GSS 30--IRS 3 (162621$-$242251)} is a class I   
source of unknown spectral type, 
15\arcsec~ to the northeast of GSS 30--IRS 1, and 11\arcsec~ southwest of 
GSS 30--IRS 2.  This source has an extended point spread function and a crescent-shaped 
morphology. 
In addition, the F160W image shows a patch of obscuration (about an arcsecond in size)
adjacent to the emission peak. 
Thus it is likely that the emission we see is scattered light from an 
embedded source.  
GSS 30--IRS 3 is also a strong radio source (LFAM 1), with emission at 
6 cm \citep{leo91}. 

\noindent
{\bf GY 30 (162625$-$242303):}
Fan-shaped nebulosity extends to the east of this apparently low-mass 
object. 

\noindent
{\bf (162704$-$243707):}  
Detected only at F160W, this source has a bipolar morphology strongly suggestive 
of a YSO outflow cavity, similar to that seen in GY 244, but fainter. 
This source was reported in Brandner et al. (2000), who imaged it at 2.2 $\mu$m. 

\noindent
{\bf WL 15 (162709$-$243718):}  
WL 15 (Elias 29) is a class I source \citep{wly89} with a molecular outflow 
\citep{bat96, sek97}. The near-IR spectrum of El 29 resembles 
that of GSS 30--IRS1; a heavily veiled, featureless spectrum with 
the Br $\gamma$ line in emission \citep{lr99}.

Simon et al. (1987) found that the best fit to their 
lunar occultation observations was a two-component model
having a central 7 mas component producing most of the emission
at 2.2~$\mu$m and a larger component 0\farcs4~ in extent, roughly centered
on the smaller one. 
Our F160W image is consistent with this. The source  
is slightly elongated and has a measured FWHM of 0\farcs33~  
(3.3 pixels in our dithered image).

\noindent
{\bf GY 244 (162717$-$242856):}
This source has a class I spectral energy distribution \citep{wly89}, and a K-band spectral 
type of M4 \citep{lr99}. 
Our images show a bipolar nebula associated with this source. 
The morphology of the nebula is characteristic of the scattered 
light seen from YSO outflow cavities \citep{wh92} and 
suggests that the SW lobe is inclined somewhat toward us. 
Faint nebulosity in the F160W image indicates that the stars near GY 244 
(GY 246, GY 247, and GY 249) also have some circumstellar envelope material 
associated with them. 

\noindent
{\bf (162724$-$244102):}
Detected only at F160W, this source has a faint bipolar morphology similar to 
that of (162704$-$243707).  
Also reported in Brandner et al. (2000). 

\noindent
{\bf YLW 15A (162726$-$244051):} 
This embedded source has a heavily veiled, 
featureless 2~$\mu$m spectrum \citep{lr99}. 
It is included in our list of binary systems (Table 3), 
on the basis of a companion located $\sim$6\arcsec~ to the northwest.

\noindent 
{\bf YLW 16A (162728$-$243934):}
We detect two non-point sources at the 
position of YLW 16A, separated by 0\farcs5~ (position angle ~270\arcdeg).
The flux ratio of the two peaks is 1.5 at 1.1~$\mu$m and  
1.1 at 1.6~$\mu$m, with a large uncertainty due to the extended
nature of the sources.
As both of these sources are extended, it unclear whether they are actually two 
embedded stars, or a single embedded star seen in scattered light; possibly 
a star/disk/envelope system. 

The appearance of the two peaks is more diffuse at 1.1~$\mu$m than at 1.6~$\mu$m, 
as would be expected if the light we detect is mainly scattered light. 
In addition, the flux ratio is greater at 1.1~$\mu$m than at 1.6~$\mu$m,
suggesting differential reddening to the two peaks. 
Lunar occultation observations at 2~$\mu$m of YLW 16A \citep{simon87} 
failed to resolve this 
source as a binary, instead showing it to be a single extended source of size 
(at K) $\sim$0\farcs5, which is the angular separation of the two peaks observed 
in our HST images. Thus it is possible that YLW 16A 
is a single star+disk+envelope, rather than an embedded 
binary system.  
 
\noindent
{\bf GY 273 (162728$-$242721):} Large sigma-shaped nebula. It is unclear 
from our images where the illuminating source is.

 
\section{Summary}\label{summ}

We have presented a NICMOS 3 survey in the F110W and F160W filters 
of the dense star-forming cores of the Ophiuchus molecular cloud. Among our 
results: 

\noindent
1. The number of sources detected at F160W is 165, 
of which 65 were detected at F110W. 
Based on a search of the literature, we estimate that 
approximately two-thirds of the sources detected at F160W 
were previously uncatalogued. Of the 108 sources located within 
the $\rm{A_V}=$40 mag. contour of the cloud, 58 were previously uncatalogued.

\noindent
2. The Ophiuchus star forming region has multiple peaks in stellar density. 
In core A we measure 
${\rm n_{\ast}= 4\times10^4\, stars\, pc^{-3}}$ within a volume 0.05-0.08 pc 
on a side, 
and in cores B and E, ${\rm n_{\ast}=0.6-5\times10^3\, stars\, pc^{-3}}$
within similar volumes.  Such densities are similar to those seen on larger 
scales in rich clusters such as the Orion Nebula cluster. 

\noindent
3. Thirteen apparent multiple systems (eight binary pairs and five triples)  
with projected separations in the  
range 0\farcs2~ to 10\arcsec~ (29 to 1450 AU) were detected. 

\noindent
4. Seven new candidate brown dwarfs
were identified from their positions in a color-magnitude  
diagram. According to our analysis of an extinction-limited sample, 
sub-stellar mass objects may account for as many as 30\% of the sources 
in the core of the Ophiuchus cluster.
 
\noindent
5. The unprecedented combination of resolution and sensitivity provided by HST 
has revealed new structures in the infrared sources in Ophiuchus. 
Bipolar structure 
is clearly seen in five objects.

\acknowledgments

We are grateful to Al Schultz at STScI for valuable advice, and to 
L. Bergeron of STScI for expert help with NICMOS data reduction.  
We thank Bruce Wilking for providing data which allowed us to reproduce 
his $\rm{A_V}$ map, and John Carpenter for sharing    
his encoded version of Wainscoat et al.'s Galaxy model.

\clearpage

\begin{figure}
\caption{Locations of observed fields
shown on an optical (DSS) image of the Ophiuchus main cloud.
Each of the 13 fields is $2\farcm45\times2\farcm45$ in size, and is composed of
a $3\times3$ mosaic of NICMOS-3 fields. Coordinates of the 13 fields are 
listed in Table 1, along with core designations (A,B,E,F) after Loren et al. 1990. 
\label{fig-geo}}
\end{figure}

\clearpage

\begin{figure}
\plotone{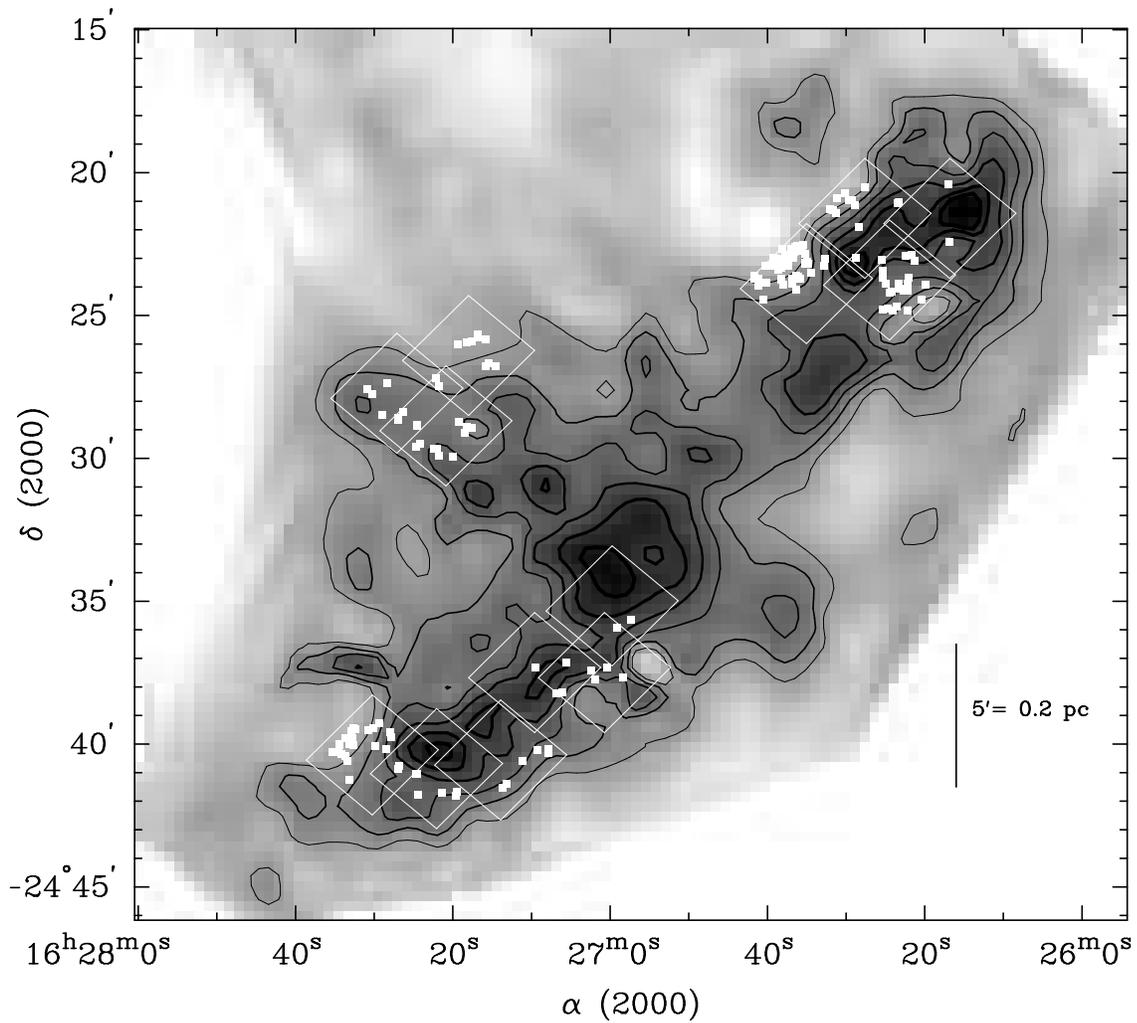}
\vskip -1.0in
\caption{Extinction map, reproduced from Wilking \& Lada (1983). Contours are plotted 
for $\rm{A_V}=$40-90 mag, at 10 mag intervals. 
Solid squares show the positions of the 165 sources detected at F160W. 
The regions covered by the 
HST survey are outlined, each box representing a $3\times3$ mosaic. 
\label{fig:avmap}}
\end{figure}

\clearpage

\begin{figure}
\plotone{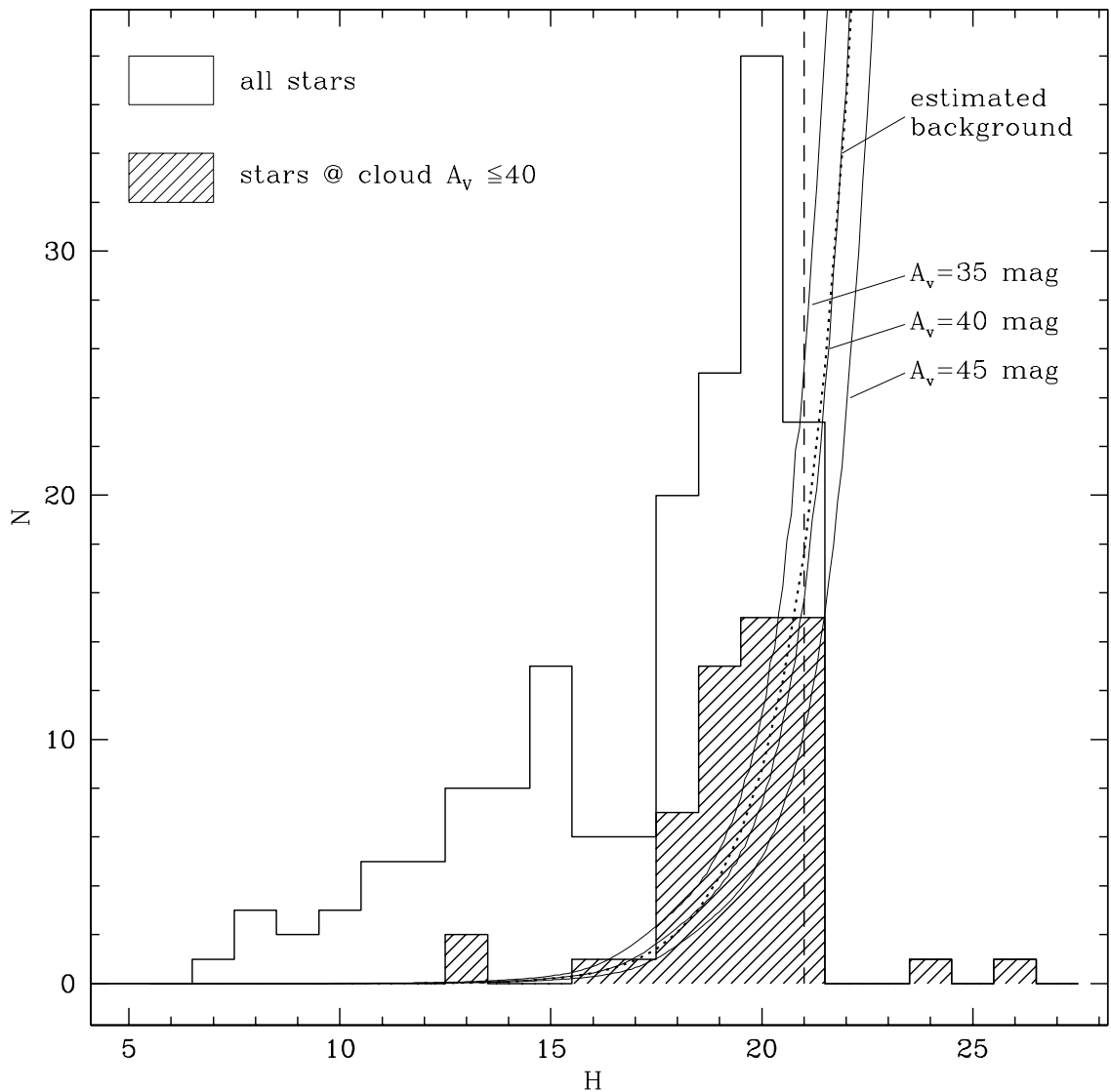}
\caption{Observed distribution of H magnitudes. The vertical dashed line 
shows the H-band completeness limit, and the heavy dotted line shows the 
predicted background, using Wainscoat's infrared Galaxy model and 
the $\rm{A_V}$ map shown in Figure~\ref{fig:avmap}. 
Model background populations assuming uniform extinction of 35, 40, and 45 
$\rm{A_V}$ are shown for comparison. 
The hatched part of the histogram represents stars whose projected locations 
on the cloud correspond to $\rm{A_V} \le$ 40 mag.
\label{fig:hlum}}
\end{figure}

\clearpage 

\begin{figure}
\plotone{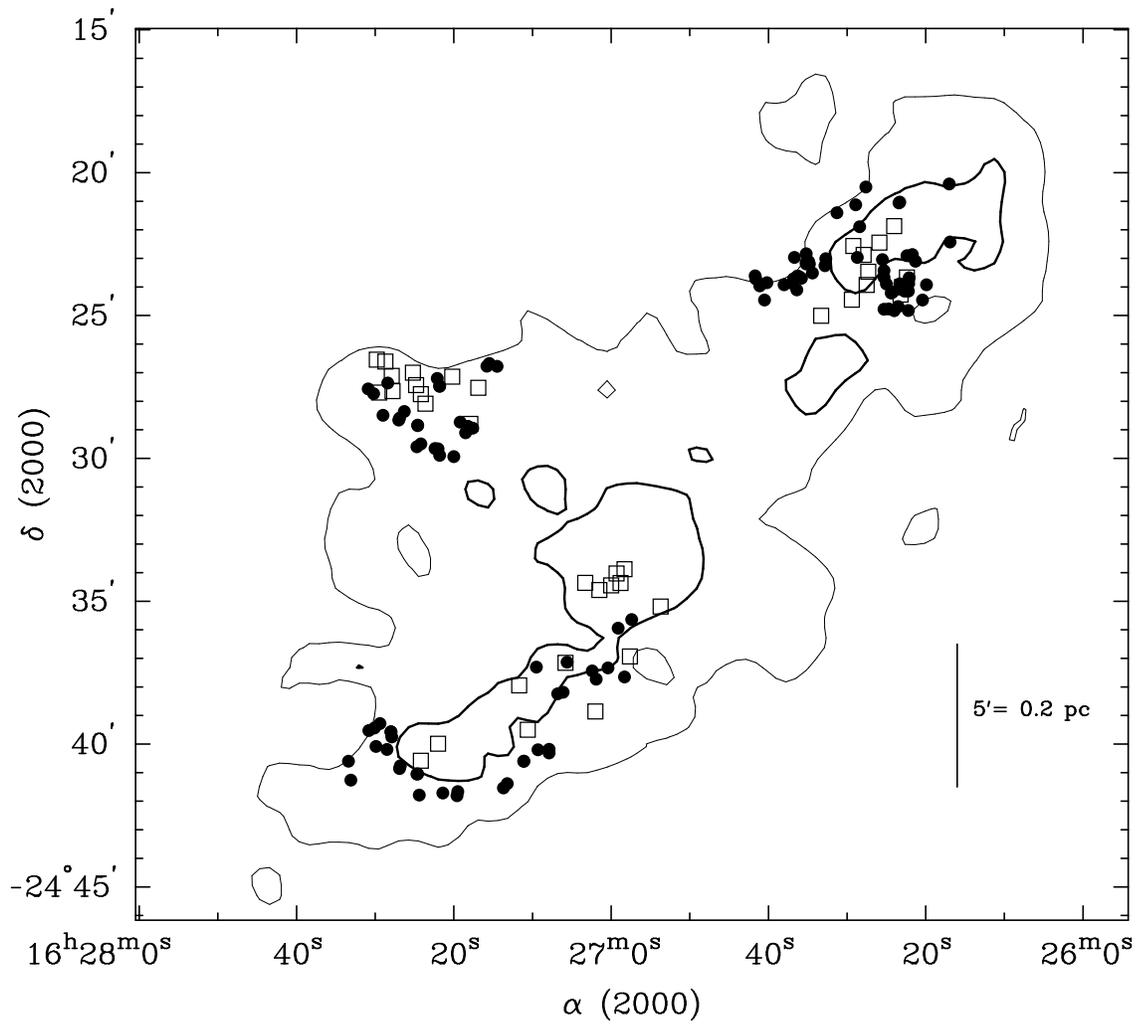}
\vskip -1.0in
\caption{Spatial distribution of the HST sample (filled circles) 
and starless dust clumps from Motte et al. (1998) (open squares). 
Contours show $\rm{A_V} =$40 and 70 mag. Note how the starless clumps 
appear to be preferentially located in regions of higher extinction.
\label{fig:space}}
\end{figure}

\clearpage

\begin{figure}
\caption{F160W band images of binary (a-h) and multiple (i-m) sources, as 
identified in Table 3. Very faint companions are encircled. The bright 
primaries in panels (l) and (m) are themselves doubles (see Table 3). 
Each image is 12\arcsec~ on a side. 
\label{fig:multis}}
\end{figure}

\clearpage

\begin{figure}
\plotone{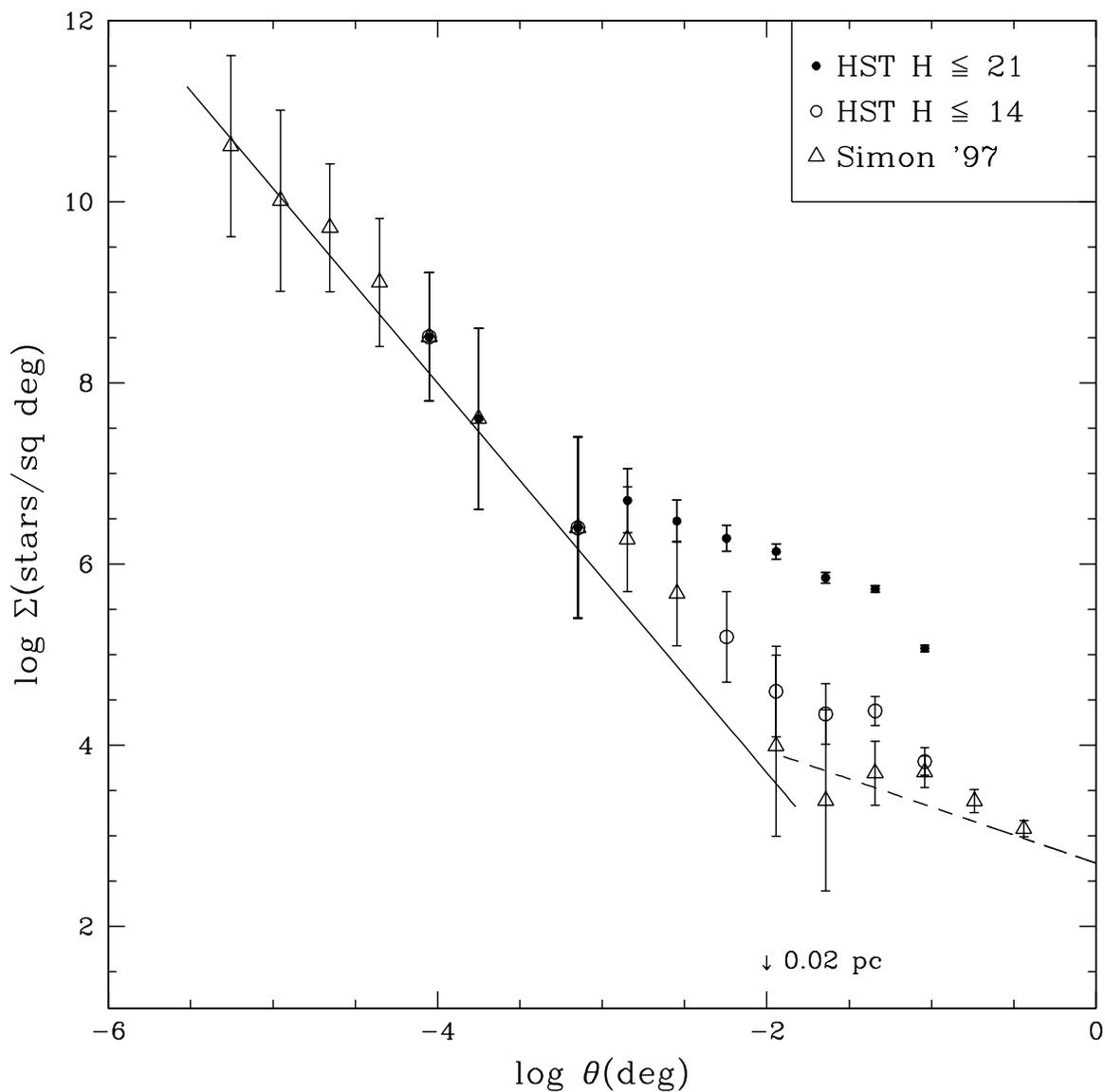}
\caption{Surface density of stars in the Ophiuchus sample, as 
a function of source separation. The break in slopes distinguishing the 
``cluster'' or ``field'' distribution from the ``binary'' distribution 
depends on the depth of the sample. Distributions are shown for this 
study, H$\le$21 (filled circles) and H$\le$14 (open circles); and for 
Simon (1997).  Larson's (1995) power--law fits are plotted as a dashed line 
for the larger separations ($\Sigma \propto \Theta^{-0.62}$) and a solid line 
for separations less than 0.04 pc ($\Sigma \propto \Theta^{-2.15}$).  
\label{fig:sd}}
\end{figure}

\clearpage

\begin{figure}
\plotone{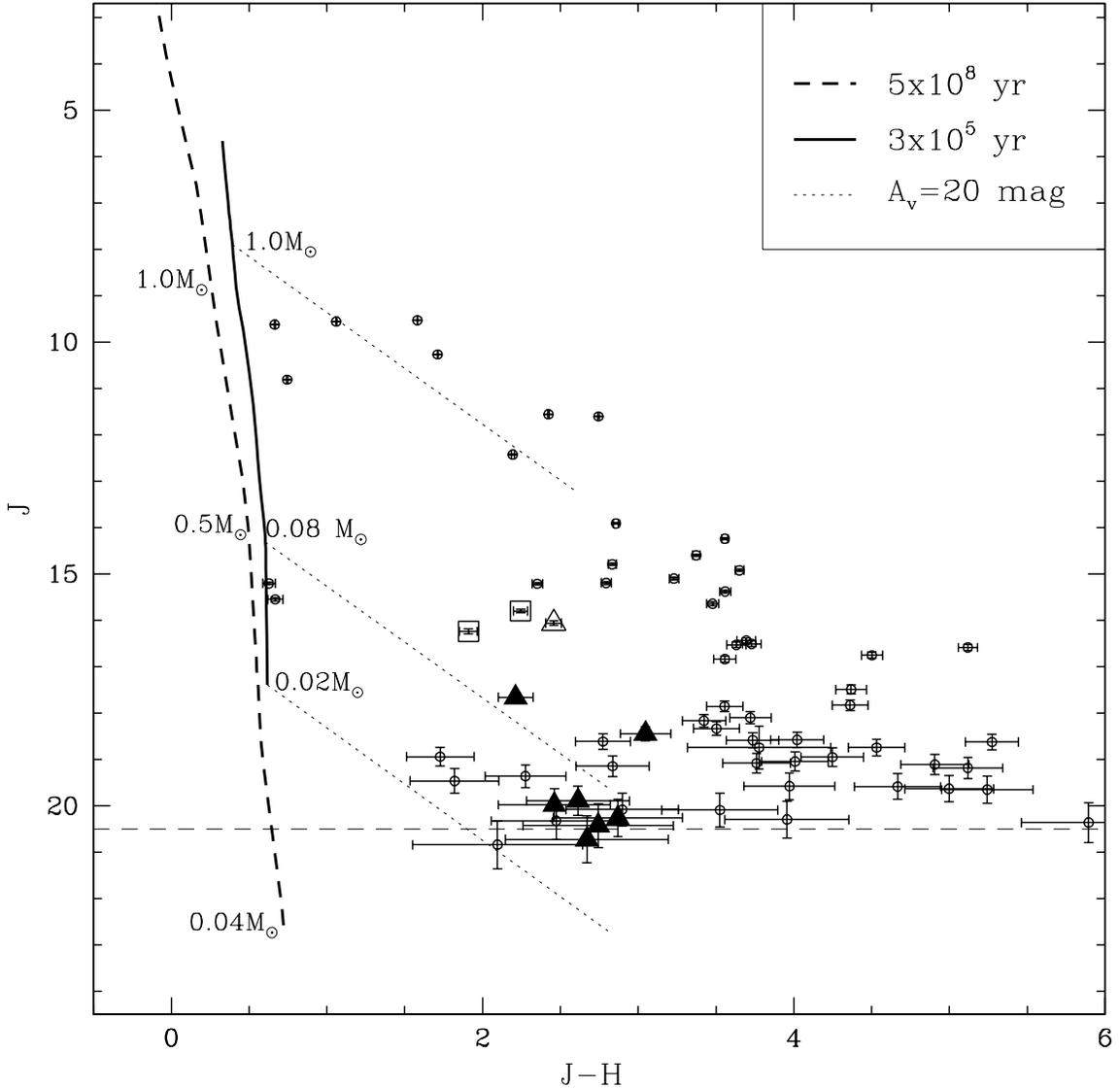}
\caption{Color-magnitude diagram for the 65 sources detected within the estimated completeness
limits in both the F110W and F160W bands. Magnitudes have been transformed from F110W
and F160W to the J, H (CIT) system using the transformation relations in Section~\ref{phot}.
Ten sources satisfying our criteria for 
brown dwarf candidacy are distinguished by their point types. 
The sources 162622$-$242409 (sub-stellar) and 162622$-$242354 
(possibly sub-stellar)  
are plotted as open squares, and the type M4 star 162659$-$243556 as 
an open triangle. 
The seven new brown dwarf candidates from this study are represented as 
solid triangles. 
Error bars represent 1$\sigma$ internal photometry errors.
\label{fig:cm}}
\end{figure}
 
\clearpage

\begin{figure}
\caption{Extended sources described in section 4. Clockwise 
from top left:
GSS 30;  GY244$+$GY247$+$GY246; WL 15; GY30;  
162724$-$244102; 162704$-$243707; YLW 16A; GY 273; YLW15A. 
Images in the F110W and F160W filters 
were combined to produce these false color images, 
except for 162724$-$244102 and 162704$-$243707, which were detected at F160W only. 
\label{fig:plate}}
\end{figure}
 
\clearpage

\begin{deluxetable}{cccc}
\tabletypesize{\small}
\singlespace
\tablecaption{Fields Observed \label{tbl-1}}
\tablewidth{0pt}
\tablehead{
\colhead{Field} & \colhead{R.A.\tablenotemark{a}} & \colhead{Dec.}   
& \colhead{Core\tablenotemark{b}}
} 
\startdata
1 & 16:26:16.0 & -24:21:30 & A \\
2 & 16:26:23.0 & -24:23:40 & A \\
3 & 16:26:25.0 & -24:21:30 & A \\
4 & 16:26:32.0 & -24:23:40 & A \\
5 & 16:26:59.0 & -24:35:10 & E \\
6 & 16:26:59.0 & -24:37:30 & E \\
7 & 16:27:08.0 & -24:37:30 & E \\
8 & 16:27:18.0 & -24:26:20 & B \\
9 & 16:27:21.0 & -24:28:50 & B \\
10 & 16:27:27.0 & -24:27:40 & B \\
11 & 16:27:14.0 & -24:40:30 & F \\
12 & 16:27:22.0 & -24:40:40 & F \\
13 & 16:27:30.0 & -24:40:10 & F \\
\enddata
\tablenotetext{a}{Coordinates are epoch J2000.}
\tablenotetext{b}{As designated by Loren et al. (1990)}. 
\end{deluxetable}

\clearpage

\begin{deluxetable}{lccccc}
\tabletypesize{\small}
\singlespace
\tablewidth{0pc}
\tablecaption{Photometry \label{tbl-2}}
\tablehead{
\colhead{ID} & \colhead{R.A.\tablenotemark{a}}   & \colhead{Dec.} 
& \colhead{m160\tablenotemark{b}}  & \colhead{m110\tablenotemark{b}}
& \colhead{cross-ref}}
\startdata
162616$-$242225 & 16:26:16.9 & -24:22:25 & 10.30 $\pm$ 0.00 & 11.13 $\pm$ 0.00  & GSS29,SKS1-8 \\
162617$-$242023 & 16:26:17.0 & -24:20:23 & 9.17 $\pm$ 0.00 & 9.86 $\pm$ 0.00  & DoAr24,SKS1-9 \\
162619$-$242355\tablenotemark{d} & 16:26:19.9 & -24:23:55 & 18.07 $\pm$ 0.11 & 21.42 $\pm$ 0.47  & \nodata \\
162620$-$242427\tablenotemark{d} & 16:26:20.4 & -24:24:27 & 17.77 $\pm$ 0.09 & 21.24 $\pm$ 0.40  & \nodata \\
162621$-$242306\tablenotemark{c} & 16:26:21.3 & -24:23:06 & 10.59 $\pm$ 0.01 & 14.20 $\pm$ 0.03  & GSS30,SKS1-12 \\
162621$-$242251 & 16:26:21.7 & -24:22:51 & 17.42 $\pm$ 0.09 & $>$21.5  & \nodata \\
162622$-$242340 & 16:26:22.1 & -24:23:40 & 17.64 $\pm$ 0.09 & $>$21.5  & \nodata \\
162622$-$242354 & 16:26:22.2 & -24:23:54 & 13.86 $\pm$ 0.01 & 16.47 $\pm$ 0.04  & GY10,SKS1-14 \\
162622$-$242449 & 16:26:22.2 & -24:24:49 & 15.66 $\pm$ 0.03 & 19.98 $\pm$ 0.21  & SKS3-12 \\
162622$-$242409 & 16:26:22.2 & -24:24:09 & 14.65 $\pm$ 0.02 & 16.94 $\pm$ 0.05  & GY11,SKS3-13 \\
162622$-$242254 & 16:26:22.4 & -24:22:54 & 12.14 $\pm$ 0.00 & 15.72 $\pm$ 0.03  & GY12 \\
162622$-$242403\tablenotemark{d} & 16:26:22.6 & -24:24:03 & 17.66 $\pm$ 0.09 & 20.85 $\pm$ 0.31  & \nodata \\
162622$-$242408\tablenotemark{d} & 16:26:22.7 & -24:24:08 & 15.79 $\pm$ 0.03 & 18.46 $\pm$ 0.10  & \nodata \\
162623$-$242404 & 16:26:23.2 & -24:24:04 & 19.70 $\pm$ 0.23 & $>$21.5  & \nodata \\
162623$-$242101 & 16:26:23.3 & -24:21:01 & 8.76 $\pm$ 0.00 & 10.55 $\pm$ 0.00  & DoAr24Ea \\
162623$-$242353\tablenotemark{d} & 16:26:23.3 & -24:23:53 & 17.90 $\pm$ 0.10 & 20.94 $\pm$ 0.34  & \nodata \\
162623$-$242103 & 16:26:23.4 & -24:21:03 & 10.47 $\pm$ 0.00 & 12.86 $\pm$ 0.00  & DoAr24Eb \\
162623$-$242441 & 16:26:23.5 & -24:24:41 & 12.23 $\pm$ 0.00 & 15.39 $\pm$ 0.02  & LFAM3,SKS1-16 \\
162624$-$242449\tablenotemark{c} & 16:26:24.0 & -24:24:49 & 8.76 $\pm$ 0.00 & 11.60 $\pm$ 0.00 & S2,SKS1-17 \\
162624$-$242410 & 16:26:24.2 & -24:24:10 & 19.96 $\pm$ 0.28 & $>$21.5   & \nodata \\
162624$-$242412 & 16:26:24.4 & -24:24:12 & 18.38 $\pm$ 0.13 & $>$21.5   & \nodata \\
162624$-$242446 & 16:26:24.7 & -24:24:46 & 16.93 $\pm$ 0.07 & 21.06 $\pm$ 0.367  & \nodata \\
162625$-$242353 & 16:26:25.0 & -24:23:53 & 19.42 $\pm$ 0.21 & $>$21.5   & \nodata \\
162625$-$242325 & 16:26:25.3 & -24:23:25 & 17.04 $\pm$ 0.07 &  & SKS3-15 \\
162625$-$242446 & 16:26:25.3 & -24:24:46 & 13.19 $\pm$ 0.01 & 17.25 $\pm$ 0.06  & GY29,SKS1-18 \\
162625$-$242339 & 16:26:25.3 & -24:23:39 & 20.64 $\pm$ 0.37 & $>$21.5  & \nodata \\
162625$-$242303\tablenotemark{c,d} & 16:26:25.5 & -24:23:03 & 14.45 $\pm$ 0.03 & 17.01 $\pm$ 0.15  & GY30,SKS3-16 \\
162627$-$242029 & 16:26:27.6 & -24:20:29 & 19.41 $\pm$ 0.20 & $>$21.5  & \nodata \\
162628$-$242153 & 16:26:28.4 & -24:21:53 & 15.30 $\pm$ 0.03 & 19.62 $\pm$ 0.45  & GY38?,SKS69 \\
162628$-$242258 & 16:26:28.7 & -24:22:58 & 18.38 $\pm$ 0.12 & $>$21.5   & \nodata \\
162628$-$242107 & 16:26:28.9 & -24:21:07 & 18.87 $\pm$ 0.16 & $>$21.5   & \nodata \\
162629$-$242056 & 16:26:29.3 & -24:20:56 & 19.89 $\pm$ 0.26 & $>$21.5   & \nodata \\
162629$-$242057 & 16:26:29.6 & -24:20:57 & 20.17 $\pm$ 0.30 & $>$21.5   & \nodata \\
162630$-$242043 & 16:26:30.2 & -24:20:43 & 20.20 $\pm$ 0.30 & $>$21.5   & \nodata \\
162631$-$242053 & 16:26:31.2 & -24:20:53 & 17.56 $\pm$ 0.08 & 21.05 $\pm$ 0.35  & \nodata \\
162631$-$242124 & 16:26:31.3 & -24:21:24 & 16.67 $\pm$ 0.05 & 20.05 $\pm$ 0.22  & \nodata \\
162631$-$242118 & 16:26:31.7 & -24:21:18 & 19.73 $\pm$ 0.25 & $>$21.5   & \nodata \\
162632$-$242116 & 16:26:32.0 & -24:21:16 & 18.59 $\pm$ 0.14 & $>$21.5   & \nodata \\
162633$-$242300 & 16:26:33.0 & -24:23:00 & 19.97 $\pm$ 0.27 & $>$21.5   & \nodata \\
162633$-$242315\tablenotemark{d} & 16:26:33.1 & -24:23:15 & 18.45 $\pm$ 0.13 & 21.74 $\pm$ 0.50  & \nodata \\
162634$-$242330\tablenotemark{c} & 16:26:34.7 & -24:23:30 & 7.73 $\pm$ 0.00 & 9.46 $\pm$ 0.00  & S1,SKS1-22 \\
162635$-$242311 & 16:26:35.1 & -24:23:11 & 20.62 $\pm$ 0.38 & $>$21.5   & \nodata \\
162635$-$242306 & 16:26:35.2 & -24:23:06 & 20.31 $\pm$ 0.32 & off field   & \nodata \\
162635$-$242250 & 16:26:35.2 & -24:22:50 & 19.47 $\pm$ 0.21 & $>$21.5  & \nodata \\
162635$-$242232 & 16:26:35.4 & -24:22:32 & 20.19 $\pm$ 0.31 & $>$21.5   & \nodata \\
162635$-$242308 & 16:26:35.5 & -24:23:08 & 19.88 $\pm$ 0.25 & off field   & \nodata \\
162635$-$242311 & 16:26:35.5 & -24:23:11 & 18.48 $\pm$ 0.13 & off field   & \nodata \\
162635$-$242341 & 16:26:35.5 & -24:23:41 & 20.19 $\pm$ 0.30 & $>$21.5   & \nodata \\
162635$-$242240 & 16:26:35.6 & -24:22:40 & 20.85 $\pm$ 0.42 & $>$21.5   & \nodata \\
162635$-$242241 & 16:26:35.6 & -24:22:41 & 19.94 $\pm$ 0.27 & $>$21.5   & \nodata \\
162635$-$242243 & 16:26:35.7 & -24:22:43 & 20.94 $\pm$ 0.44 & $>$21.5   & \nodata \\
162635$-$242338 & 16:26:35.8 & -24:23:38 & 21.20 $\pm$ 0.50 & $>$21.5   & \nodata \\
162636$-$242336 & 16:26:36.0 & -24:23:36 & 17.46 $\pm$ 0.08 & 20.28 $\pm$ 0.24  & \nodata \\
162636$-$242406 & 16:26:36.1 & -24:24:06 & 15.95 $\pm$ 0.04 & 20.51 $\pm$ 0.28  & GY76,SKS3-19 \\
162636$-$242232 & 16:26:36.2 & -24:22:32 & 20.95 $\pm$ 0.45 & $>$21.5   & \nodata \\
162636$-$242236 & 16:26:36.5 & -24:22:36 & 19.64 $\pm$ 0.23 & $>$21.5 & \nodata \\
162636$-$242342 & 16:26:36.5 & -24:23:42 & 19.41 $\pm$ 0.20 & $>$21.5   & \nodata \\
162636$-$242245 & 16:26:36.6 & -24:22:45 & 19.98 $\pm$ 0.27 & $>$21.5 & \nodata \\
162636$-$242353 & 16:26:36.6 & -24:23:53 & 16.19 $\pm$ 0.04 & 19.48 $\pm$ 0.17  & SKS3-21 \\
162636$-$242258 & 16:26:36.7 & -24:22:58 & 19.89 $\pm$ 0.28 & $>$21.5   & \nodata \\
162637$-$242247 & 16:26:37.0 & -24:22:47 & 18.03 $\pm$ 0.10 & 20.39 $\pm$ 0.26  & \nodata \\
162637$-$242239 & 16:26:37.0 & -24:22:39 & 17.59 $\pm$ 0.08 & 19.83 $\pm$ 0.19  & \nodata \\
162637$-$242256 & 16:26:37.5 & -24:22:56 & 20.51 $\pm$ 0.38 & $>$21.5   & \nodata \\
162637$-$242316 & 16:26:37.6 & -24:23:16 & 18.24 $\pm$ 0.12 & 21.31 $\pm$ 0.40  & \nodata \\
162637$-$242253 & 16:26:37.7 & -24:22:53 & 19.37 $\pm$ 0.21 & $>$21.5   & \nodata \\
162637$-$242355 & 16:26:37.7 & -24:23:55 & 19.88 $\pm$ 0.28 & off field   & \nodata \\
162637$-$242302 & 16:26:37.8 & -24:23:02 & 12.68 $\pm$ 0.00 & 15.82 $\pm$ 0.03  & GY81,SKS1-23 \\
162638$-$242241 & 16:26:38.1 & -24:22:41 & 16.59 $\pm$ 0.05 & off field   & \nodata \\
162638$-$242314 & 16:26:38.4 & -24:23:14 & 20.44 $\pm$ 0.34 & $>$21.5   & \nodata \\
162638$-$242345 & 16:26:38.5 & -24:23:45 & 18.55 $\pm$ 0.14 & $>$21.5   & \nodata \\
162638$-$242313 & 16:26:38.5 & -24:23:13 & 20.67 $\pm$ 0.39 & $>$21.5   & \nodata \\
162638$-$242342 & 16:26:38.5 & -24:23:42 & 20.25 $\pm$ 0.33 & $>$21.5   & \nodata \\
162638$-$242320 & 16:26:38.5 & -24:23:20 & 23.81 $\pm$ 2.90 & $>$21.5   & \nodata \\
162638$-$242300 & 16:26:38.6 & -24:23:00 & 20.73 $\pm$ 0.39 & $>$21.5   & \nodata \\
162638$-$242317 & 16:26:38.6 & -24:23:17 & 20.58 $\pm$ 0.36 & $>$21.5   & \nodata \\
162638$-$242312 & 16:26:38.7 & -24:23:12 & 19.15 $\pm$ 0.18 & 21.86 $\pm$ 0.51  & \nodata \\
162638$-$242324 & 16:26:38.8 & -24:23:24 & 13.15 $\pm$ 0.01 & 15.84 $\pm$ 0.03  & GY84,SKS24 \\
162639$-$242258 & 16:26:39.4 & -24:22:58 & 25.30 $\pm$ 7.42 & off field & \nodata \\
162639$-$242315 & 16:26:39.4 & -24:23:15 & 19.70 $\pm$ 0.23 & $>$21.5   & \nodata \\
162640$-$242351 & 16:26:40.2 & -24:23:51 & 19.80 $\pm$ 0.25 & $>$21.5   & \nodata \\
162640$-$242315 & 16:26:40.3 & -24:23:15 & 16.44 $\pm$ 0.05 & off field & \nodata \\
162640$-$242427 & 16:26:40.5 & -24:24:27 & 18.25 $\pm$ 0.12 & $>$21.5   & \nodata \\
162641$-$242342 & 16:26:41.1 & -24:23:42 & 19.22 $\pm$ 0.18 & $>$21.5   & \nodata \\
162641$-$242357 & 16:26:41.1 & -24:23:57 & 20.25 $\pm$ 0.30 & $>$21.5   & \nodata \\
162641$-$242343 & 16:26:41.6 & -24:23:43 & 19.29 $\pm$ 0.19 & $>$21.5   & \nodata \\
162641$-$242336 & 16:26:41.7 & -24:23:36 & 18.57 $\pm$ 0.13 & $>$21.5   & \nodata \\
162657$-$243538 & 16:26:57.4 & -24:35:38 & 15.18 $\pm$ 0.02 & 19.45 $\pm$ 0.166  & SKS3-23 \\
162658$-$243739 & 16:26:58.3 & -24:37:39 & 18.05 $\pm$ 0.10 & $>$21.5   & CRBR51,SKS3-24 \\
162659$-$243556 & 16:26:59.1 & -24:35:56 & 13.91 $\pm$ 0.01 & 16.75 $\pm$ 0.048  & SKS3-25 \\
162700$-$243719 & 16:27:00.4 & -24:37:19 & 20.89 $\pm$ 0.41 & $>$21.5   & \nodata \\
162701$-$243743 & 16:27:01.9 & -24:37:43 & 20.04 $\pm$ 0.29 & $>$21.5   & \nodata \\
162702$-$243726 & 16:27:02.4 & -24:37:26 & 10.93 $\pm$ 0.00 & 14.80 $\pm$ 0.019  & WL16,SKS1-25 \\
162704$-$243707\tablenotemark{c} & 16:27:04.6 & -24:37:07 & 17.59 $\pm$ 0.18 & $>$21.5   & \nodata \\
162706$-$243811 & 16:27:06.1 & -24:38:11 & 14.90 $\pm$ 0.02 & 15.83 $\pm$ 0.03  & GY201 \\
162706$-$243814 & 16:27:06.8 & -24:38:14 & 13.77 $\pm$ 0.01 & 18.64 $\pm$ 0.11  & WL17,SKS3-27 \\
162707$-$244011 & 16:27:07.9 & -24:40:11 & 19.95 $\pm$ 0.27 & $>$21.5   & \nodata \\
162707$-$244018 & 16:27:07.9 & -24:40:18 & 16.70 $\pm$ 0.05 & 21.28 $\pm$ 0.39  & \nodata \\
162709$-$244011 & 16:27:09.3 & -24:40:11 & 13.58 $\pm$ 0.01 & 17.58 $\pm$ 0.07  & \nodata \\
162709$-$243718\tablenotemark{c} & 16:27:09.5 & -24:37:18 & 11.05 $\pm$ 0.00 & 16.51 $\pm$ 0.06 & GY214,WL 15, El 29, SKS1-28 \\
162711$-$244036 & 16:27:11.1 & -24:40:36 & 13.03 $\pm$ 0.01 & 17.15 $\pm$ 0.05  & \nodata \\
162711$-$243831 & 16:27:11.4 & -24:38:31 & 15.29 $\pm$ 0.03 & 21.10 $\pm$ 0.04  & \nodata \\
162713$-$244123 & 16:27:13.2 & -24:41:23 & 12.53 $\pm$ 0.00 & 17.49 $\pm$ 0.06  & \nodata \\
162713$-$244132 & 16:27:13.7 & -24:41:32 & 19.54 $\pm$ 0.21 & $>$21.5 $\pm$ \nodata  & \nodata \\
162714$-$242646 & 16:27:14.5 & -24:26:46 & 15.25 $\pm$ 0.03 & 20.52 $\pm$ 0.27  & GY236,SKS3-34 \\
162715$-$242640 & 16:27:15.5 & -24:26:40 & 13.42 $\pm$ 0.01 & 18.28 $\pm$ 0.09  & GY239,SKS3-36 \\
162715$-$242646 & 16:27:15.8 & -24:26:46 & 20.12 $\pm$ 0.28 & $>$21.5   & \nodata \\
162715$-$242550 & 16:27:15.8 & -24:25:50 & 19.99 $\pm$ 0.27 & $>$21.5   & \nodata \\
162716$-$242540 & 16:27:16.8 & -24:25:40 & 19.14 $\pm$ 0.18 & $>$21.5   & \nodata \\
162716$-$242546 & 16:27:16.9 & -24:25:46 & 20.35 $\pm$ 0.33 & $>$21.5   & \nodata \\
162717$-$242856\tablenotemark{c} & 16:27:17.6 & -24:28:56 & 13.63 $\pm$ 0.02 & 18.04 $\pm$ 0.16  & GY244,SKS1-32 \\
162717$-$242554 & 16:27:17.6 & -24:25:54 & 18.55 $\pm$ 0.14 & $>$21.5   & \nodata \\
162717$-$242554 & 16:27:17.6 & -24:25:54 & 18.61 $\pm$ 0.14 & $>$21.5   & \nodata \\
162718$-$242853 & 16:27:18.2 & -24:28:53 & 14.73 $\pm$ 0.02 & 20.59 $\pm$ 0.29  & WL5,SKS1-33 \\
162718$-$242555 & 16:27:18.2 & -24:25:55 & 18.89 $\pm$ 0.16 & $>$21.5   & \nodata \\
162718$-$242906\tablenotemark{c} & 16:27:18.5 & -24:29:06 & 11.02 $\pm$ 0.00 & 14.26 $\pm$ 0.01  & WL4,SKS1-34 \\
162719$-$242844\tablenotemark{c} & 16:27:19.2 & -24:28:44 & 14.41 $\pm$ 0.02 & 19.15 $\pm$ 0.28  & WL3,SKS3-38 \\
162719$-$242601 & 16:27:19.4 & -24:26:01 & 18.67 $\pm$ 0.14 & $>$21.5   & \nodata \\
162719$-$244139 & 16:27:19.5 & -24:41:39 & 8.70 $\pm$ 0.00 & 9.79 $\pm$ 0.00  & SR12,SKS1-35 \\
162719$-$244148 & 16:27:19.6 & -24:41:48 & 15.21 $\pm$ 0.03 & 16.20 $\pm$ 0.03  & \nodata \\
162720$-$242956 & 16:27:20.0 & -24:29:56 & 20.03 $\pm$ 0.27 & $>$21.5   & \nodata \\
162721$-$244142 & 16:27:21.4 & -24:41:42 & 11.48 $\pm$ 0.00 & 15.18 $\pm$ 0.02  & YLW13b,SKS1-36 \\
162721$-$242728 & 16:27:21.8 & -24:27:28 & 18.71 $\pm$ 0.14 & $>$21.5   & \nodata \\
162721$-$242953 & 16:27:21.8 & -24:29:53 & 14.38 $\pm$ 0.02 & 20.09 $\pm$ 0.22  & GY254,SKS3-40 \\
162722$-$242940 & 16:27:22.0 & -24:29:40 & 15.37 $\pm$ 0.03 & 19.94 $\pm$ 0.21  & GY256,SKS3-41 \\
162722$-$242712 & 16:27:22.1 & -24:27:12 & 19.83 $\pm$ 0.25 & $>$21.5   & \nodata \\
162722$-$242939 & 16:27:22.4 & -24:29:39 & 19.89 $\pm$ 0.26 & $>$21.5   & \nodata \\
162724$-$242929 & 16:27:24.2 & -24:29:29 & 15.03 $\pm$ 0.02 & 19.84 $\pm$ 0.20  & GY257,SKS3-43 \\
162724$-$244147 & 16:27:24.4 & -24:41:47 & 15.07 $\pm$ 0.02 & 19.00 $\pm$ 0.13  & GY258,SKS3-42 \\
162724$-$242851 & 16:27:24.6 & -24:28:51 & 20.54 $\pm$ 0.36 & $>$21.5   & \nodata \\
162724$-$242850 & 16:27:24.6 & -24:28:50 & 20.00 $\pm$ 0.27 & $>$21.5   & \nodata \\
162724$-$244102 & 16:27:24.6 & -24:41:02 & 18.60 $\pm$ 0.14 & $>$21.5   & \nodata \\
162724$-$242935 & 16:27:24.7 & -24:29:35 & 14.70 $\pm$ 0.02 & 18.93 $\pm$ 0.13  & GY259,SKS3-45 \\
162724$-$244103 & 16:27:24.7 & -24:41:03 & 18.55 $\pm$ 0.14 & $>$21.5   & CRBR85,SKS3-44 \\
162726$-$242821 & 16:27:26.3 & -24:28:21 & 19.11 $\pm$ 0.18 & $>$21.5   & \nodata \\
162726$-$244045\tablenotemark{c} & 16:27:26.8 & -24:40:45 & 14.75 $\pm$ 0.02 & 18.54 $\pm$ 0.14  & GY263,SKS3-48 \\
162726$-$244051\tablenotemark{c} & 16:27:26.9 & -24:40:51 & 13.05 $\pm$ 0.01 & 18.22 $\pm$ 0.16  & YLW15A,SKS3-49 \\
162726$-$242835 & 16:27:26.9 & -24:28:35 & 20.83 $\pm$ 0.40 & $>$21.5   & \nodata \\
162727$-$242839 & 16:27:27.0 & -24:28:39 & 20.76 $\pm$ 0.39 & $>$21.5   & \nodata \\
162727$-$244048 & 16:27:27.1 & -24:40:48 & 18.21 $\pm$ 0.12 & $>$21.5   & \nodata \\
162727$-$243944 & 16:27:27.9 & -24:39:44 & 18.82 $\pm$ 0.15 & $>$21.5   & \nodata \\
162728$-$243934A:B\tablenotemark{c} & 16:27:28.0 & -24:39:34 & 12.62 $\pm$ 0.02 & 17.15 $\pm$ 0.18  & YLW16A,SKS3-51 \\
162728$-$242721 & 16:27:28.4 & -24:27:21 & 12.44 $\pm$ 0.00 & 16.30 $\pm$ 0.03  & GY 273,VSSG18,SKS1-39 \\
162728$-$244011 & 16:27:28.5 & -24:40:11 & 18.02 $\pm$ 0.10 & $>$21.5   & \nodata \\
162729$-$244005 & 16:27:29.9 & -24:40:05 & 20.69 $\pm$ 0.37 & $>$21.5   & \nodata \\
162729$-$242829 & 16:27:29.0 & -24:28:29 & 19.25 $\pm$ 0.19 & $>$21.5   & \nodata \\
162729$-$243917\tablenotemark{c} & 16:27:29.4 & -24:39:17 & 14.29 $\pm$ 0.02 & 18.28 $\pm$ 0.11  & GY274,SKS3-54 \\
162730$-$243926 & 16:27:30.1 & -24:39:26 & 19.80 $\pm$ 0.25 & $>$21.5   & \nodata \\
162730$-$242744 & 16:27:30.2 & -24:27:44 & 11.53 $\pm$ 0.00 & 15.53 $\pm$ 0.02  & VSSG17,SKS1-40 \\
162730$-$243931 & 16:27:30.8 & -24:39:31 & 20.25 $\pm$ 0.32 & $>$21.5   & \nodata \\
162730$-$242734 & 16:27:30.9 & -24:27:34 & 19.22 $\pm$ 0.19 & $>$21.5   & \nodata \\
162732$-$243928 & 16:27:32.4 & -24:39:28 & 20.62 $\pm$ 0.37 & $>$21.5   & \nodata \\
162732$-$243926 & 16:27:32.6 & -24:39:26 & 18.84 $\pm$ 0.15 & $>$21.5   & \nodata \\
162732$-$244002 & 16:27:32.6 & -24:40:02 & 19.16 $\pm$ 0.18 & $>$21.5   & \nodata \\
162732$-$243947 & 16:27:32.8 & -24:39:47 & 17.93 $\pm$ 0.10 & $>$21.5   & \nodata \\
162732$-$243932 & 16:27:32.9 & -24:39:32 & 20.59 $\pm$ 0.36 & $>$21.5   & \nodata \\
162733$-$244115 & 16:27:33.1 & -24:41:15 & 9.35 $\pm$ 0.00 & 11.93 $\pm$ 0.00  & GY292,SKS1-43 \\
162733$-$244000 & 16:27:33.1 & -24:40:00 & 18.25 $\pm$ 0.12 & $>$21.5   & \nodata \\
162733$-$244036 & 16:27:33.4 & -24:40:36 & 16.42 $\pm$ 0.05 & $>$21.5   & CRBR91 \\
162733$-$244025 & 16:27:33.6 & -24:40:25 & 20.87 $\pm$ 0.42 & $>$21.5   & \nodata \\
162733$-$243947 & 16:27:33.6 & -24:39:47 & 20.23 $\pm$ 0.31 & $>$21.5   & \nodata \\
162734$-$244021 & 16:27:34.1 & -24:40:21 & 20.70 $\pm$ 0.39 & $>$21.5   & \nodata \\
162734$-$244001 & 16:27:34.4 & -24:40:01 & 19.25 $\pm$ 0.20 & $>$21.5   & \nodata \\
162734$-$244017 & 16:27:34.6 & -24:40:17 & 20.57 $\pm$ 0.35 & $>$21.5   & \nodata \\
162735$-$244017 & 16:27:34.9 & -24:40:17 & 18.33 $\pm$ 0.12 & $>$21.5   & \nodata \\
162735$-$244016 & 16:27:35.3 & -24:40:16 & 20.82 $\pm$ 0.40 & $>$21.5   & \nodata \\
\enddata
 
\tablenotetext{a}{Coordinates are epoch J2000.}
\tablenotetext{b}{Quoted errors are statistical photometric errors and 
do not take into account uncertainty in the flux calibration, which may be higher.}
\tablenotetext{c}{Extended source; photometry reported for flux within a 3\arcsec~ aperture.}
\tablenotetext{d}{Brown dwarf candidate} 
\end{deluxetable}
 
\doublespace

\clearpage
 
\begin{deluxetable}{llccccll}
\tabletypesize{\small}
\singlespace
\tablewidth{0pc}
\tablecaption{Multiple Systems \label{tbl-3}}
\tablehead{
\colhead{$\rm{ID_A}$} & \colhead{$\rm{ID_B}$} & \colhead{$\rm{m160_A}$} & \colhead{$\rm{m160_B}$}
& \colhead{separation(\arcsec)}  & \colhead{P.A.(\arcdeg)}
& \colhead{$\rm{cross-ref_A}$}  
& \colhead{fig\tablenotemark{a}}}
\startdata
                &                 &      & Binaries & &      &      &     \\
162625$-$242446 & 162624$-$242446 & 13.2 & 16.9 & 9.0 & 274  & GY29 & (a)  \\
162634$-$242330A\tablenotemark{\dag} & 162634$-$242330B & 7.7\tablenotemark{b} & -- &      &      & S1 &   \\
162636$-$242336 & 162636$-$242342 & 17.5 & 19.4 & 9.5 & 127  &  & (b)  \\
162707$-$244018 & 162707$-$244011 & 16.7 & 19.9 & 7.7 & 354  &  & (c)  \\
162717$-$242856 & 162718$-$242853 & 13.6 & 14.7 & 9.7 &  65  & GY244 & (d)  \\
162722$-$242940 & 162722$-$242939 & 15.4 & 19.9 & 6.0 &  84  & GY256 & (e)  \\
162724$-$242935 & 162724$-$242929 & 14.7 & 15.0 & 9.5 & 314  & GY259 & (f)  \\
162726$-$244051 & 162726$-$244045 & 13.0 & 14.7 & 5.9 & 323 & YLW15A  & (g)  \\
162728$-$243934A & 162728$-$243934B & 15.9\tablenotemark{b} & -- & 0.6 & 260 & YLW16A  & (h)  \\
                &                 &      & Triples  & &    &   &      \\
162622$-$242409 & 162622$-$242408 & 14.6 & 15.8 & 7.5 & 88 & GY11 & (i)  \\
                & 162622$-$242403 &      & 17.7 & 7.7 & 42 &   &      \\
162622$-$242408 & 162622$-$242403 & 15.8 & 17.7 & 4.9 & 340 &  & (j)  \\
                & 162623$-$242404 &      & 19.7 & 8.6 & 57 &   &       \\
162636$-$242336 & 162635$-$242338 & 17.5 & 20.2 & 3.4 & 228 &  &   (k)  \\
                & 162635$-$242341 &      & 20.2 & 9.0 & 234 &   &       \\
162715$-$242640A & 162715$-$242640B & 13.4\tablenotemark{b} & & 0.3 & 48 & GY239 &  (l) \\ 
                 & 162715$-$242646 &  & 20.1 & 7.3 & 145 & &  \\
162719$-$244139A\tablenotemark{\ddag} & 162719$-$244139B & 8.7\tablenotemark{b} &  & 0.3 & 96 & SR 12  &  (m) \\
                & 162719$-$244148 &      & 15.2 & 8.8 & 166 &  &      \\
                & 162727$-$244048 &      & 18.2 & 4.4 & 3 &   &       \\
\enddata
\tablenotetext{a}{Refers to panel in figure 5.}
\tablenotetext{b}{Composite magnitude.}
\tablenotetext{\dag}{S1 (162634$-$242330), a known binary of separation 
0\farcs02 \citep{simon95}, was covered by our survey but was unresolved in 
our images. It is included in this table for completeness.}
\tablenotetext{\ddag}{We detect a companion 8\farcs8~ from SR 12 (162719$-$244139A:B), 
itself a binary of separation 0\farcs28.} 
\end{deluxetable}


\clearpage
 









\clearpage


\begin{thebibliography}{}
\bibitem[Adams \& Myers (2001)]{adams2001} Adams, F.C. \& Myers, P.C. 2001, \apj, 553, 744
\bibitem[Barsony et al.(1989)]{bar89} Barsony, M., Burton, M. G., Russell, A. P. G., Carlstrom, J. E., and Garden, R. 1989, \apj, 346, L93 
\bibitem[Barsony et al.(1997)]{bklt} Barsony, M., Kenyon, S. J., Lada, E. A., and Teuben, P. J. 1997, \apjs, 112, 109
\bibitem[Bontemps, et al.(1996)]{bat96} Bontemps, S., Andr\'e, P., Terebey, S. and Cabrit, S. 1996, \aap, 311, 858  
\bibitem[Carpenter (2000)]{carp00} Carpenter, J. M. 2000, \aj, 120, 3139
\bibitem[Castelaz et al.(1985)]{cas85} Castelaz, M. W., Hackwell, J. A., Grasdalen, G. L., Gehrz, R. D. and Gullixson, C. 1985, \apj, 290, 261 
\bibitem[Chelli et al.(1988)]{chelli88} Chelli, A., Zinnecker, H., Carrasco, L. 
 Cruz-Gonz\'{a}les, I., and Perrier, C. 1988, \aap, 207, 46
\bibitem[Chrysostomou et al.(1997)]{chrys97} Chrysostomou, A., M\'enard, F., Gledhill, T. M., Clark, S., Hough, J. H., McCall, A. and Tamura, M. 1997, \mnras 285, 750 
\bibitem[Chrysostomou et al.(1996)]{chrys96} Chrysostomou, A., Clark, S. G., Hough, J. H., Gledhill, T. M.,McCall, A. and Tamura, M. 1997, \mnras 278, 449 
\bibitem[Cohen et al.(1982)]{coh81} Cohen, J.G., Elias, J.H., Frogel, J.A., and Persson, S.E. 1981, \apj, 249, 502  
\bibitem[Comeron et al.(1993)]{com93} Comeron, F., Rieke, G. H., Burrows, A., and Rieke, M. J. 1993, \apj, 416, 185  
\bibitem[de Zeeuw et al.(1999)]{deZ99} de Zeeuw, P. T., Hoogerwerf, R., de Bruijne, J. H. J., Brown, A. G. A., and Blaauw, A. 1999, \aj, 117, 354 
\bibitem[Duquennoy \& Mayor (1991)]{dm91} Duquennoy, A. \& Mayor, M. 1991, \aap, 248, 485 
\bibitem[Elias (1978)]{el78} Elias, J. H. 1978, \apj, 224, 453 
\bibitem[Fazio et al.(1976)]{faz76} Fazio, G. G., Wright, E. L., Zeilik, and Low, F.J. 1976, \apj, 206, L165 
\bibitem[Fruchter \& Hook (1997)]{fru97} Fruchter, A. S., and Hook, R. N. 1997, Proc. SPIE, 3164 
\bibitem[Ghez, Neugebauer, \& Matthews (1993)]{ghez93} Ghez, A. M., Neugebauer, G., and Matthews, K. 1993, \aj, 106, 2005
\bibitem[Grasdalen, Strom, \& Strom (1973)]{gss} Grasdalen, G. L., Strom, K. M., and strom, S. E. 1973, \apj, 184, L53 
\bibitem[Greene \& Young (1992)]{gy} Greene, T. P. and Young, E. T. 1992, \apj, 395, 516  
\bibitem[Greene \& Meyer (1995)]{gm95} Greene, T. P. and Meyer, M. R. 1995, \apj, 450, 233  
\bibitem[Hillenbrand \& Hartmann (1998)]{hill} Hillenbrand, L. A. \& Hartmann, L. W. 1998, \apj, 492, 540
\bibitem[Kirkpatrick et al.(1999)]{jdk99} Kirkpatrick, J. D., Ried, I. N., Leibert, J., Cutri, R. M., Nelson, B., Beichman, C. A., Dahn, C. C., Monet, D. G., Gizis, J. E., and Skrutskie, M. F. 1999, \apj, 519, 802 
\bibitem[Klessen et al (2000)]{klessen} Klessen, R. and so forth 
\bibitem[Leous et al.(1991)]{leo91} Leous, J. A., Feigelson, E. D., Andr\'e, P., and Montmerle, T. 1991, \apj, 379, 683  
\bibitem[Loren \& Wootten (1986)]{lor86} Loren, R. B., and Wootten, A. 1986, \apj, 306, 142
\bibitem[Loren, Wootten \& Wilking (1990)]{lww90} Loren, R. B., Wootten, A., and Wilking, 
    B. A. 1990, \apj, 365, 269 
\bibitem[Luhman \& Rieke (1999)]{lr99} Luhman, K. L. and Rieke, G. H. 1999, \apj, 525, 440 
\bibitem[Luhman (1999)]{luh} Luhman, K. L. 1999, private communication
\bibitem[Meyer et al. (2000)]{meyer} Meyer, M.R., Adams, F.C., Hillenbrand, L.A., Carpenter, J.M., and Larson, R.B. 2000, Protostars and Planets IV (University of Arizona Press, eds 
Mannings, V., Boss, A.P., Russell, S.S.), p. 121
\bibitem[Myers (2000)]{myers2000} Myers, P.C. 2000, \apj, 530, L119 
\bibitem[Motte et al.(1998)]{mot98} Motte, F., Andr\'{e}, P. and Neri, R. 1998, \aap, 336,150 
\bibitem[Nakajima et al.(1998)]{nak98} Nakajima, Y., Tachihara, K., Hanawa, T., and Nakano, M. 1998, \apj, 497, 721 
\bibitem[Reipurth \& Zinnecker (1993)]{reizen93} Reipurth, B. \& Zinnecker, H. 1993, \aa, 278, 81
\bibitem[Rieke, Ashok, \& Boyle (1989)]{rab89} Rieke, G. H., Ashok, N. M., and Boyle, R. P. 1989, \apj, 339, L71
\bibitem[Rieke (1999)]{rieke99} Rieke, M. 1999, private communication 
\bibitem[Sekimoto et al.(1997)]{sek97} Sekimoto, Y., Tatematsu, K., Umemoto, T., Koyama, K., Tsuboi, Y., Hirano, N., and Yamamoto, S. 1997, \apj, 489, L63  
\bibitem[Simon et al.(1987)]{simon87} Simon, M., Howell, R. R., Longmore, A. J., Wilking, 
    B. A., Peterson, D. M., and Chen, W.-P. 1987, \apj, 320, 344
\bibitem[Simon et al.(1995)]{simon95} Simon, M., Ghez, A. M., Leinert, Ch., 
    Cassar, L., Chen, W. P., Howell, R. R., Jameson, R. F., Matthews, K., 
    Neugebauer, G., and Richichi, A. 1995, \apj, 443, 625
\bibitem[Simon (1997)]{simon97} Simon, M. 1997, \apj, 482, L81  
\bibitem[Strom, Kepner \& Strom (1995)]{sks} Strom, K. M., Kepner, J., and Strom, S. E. 1995, \apj, 438, 813
\bibitem[Tamura et al.(1991)]{tam91} Tamura, M., Gatley, I., Joyce, R. R., Ueno, M., Suto, H., 
    and Sekiguchi, M. 1991, \apj, 378, 611 
\bibitem[Vrba et al.(1973)]{vrb73} Vrba, F. J., Strom, K. M., Strom, S. E., and Grasdalen, G. L. 1975, \apj, 197, 77 
\bibitem[Wainscoat et al.(1992)]{wain92} Wainscoat, R. J., Cohen, M., Volk, K.,Walker, H. J., 
    and Schwartz, D. E. 1992, \apjs, 83, 111
\bibitem[Weintraub et al.(1993)]{wein93} Weintraub, D. A., Kastner, J. H., Griffith, L. L., 
    and Campins, H. 1993, \aj, 105, 271 
\bibitem[Whitney \& Hartmann (1992)]{wh92} Whitney, B. A. \& Hartmann, L. 1992, \apj, 395, 529
\bibitem[Wilking \& Lada (1983)]{wil83} Wilking, B. A. and Lada, C. J. 1983, \apj, 274, 698
\bibitem[Wilking, Lada, \& Young (1989)]{wly89} Wilking,  B. A., Lada, C. J. and Young, 
    E. T. 1989, \apj, 340, 823  
\bibitem[Wilking, Greene, \& Meyer (1999)]{wgm99} Wilking, B. A., Greene, T. P., and Meyer, M. R. 1999, 
    \aj, 117, 469  
\bibitem[Wilking (1999)]{wil2} Wilking, B. A. 1999, private communication
\bibitem[Zhang et al.(1997)]{qz97} Zhang, Q., Wootten, A. and Ho, P. T. P. 1997, \apj, 475, 713   
\end{thebibliography}
\end{document}